\documentclass[epj,twocolumn]{webofc}

\usepackage[varg]{txfonts}   
\usepackage{textcomp}
\usepackage{amsmath}
\usepackage{mathrsfs}
\usepackage{amssymb}
\usepackage{slashed}
\usepackage[english]{babel}
\usepackage{graphicx}
\newcommand{\de}{\,\mathrm{d}}

\newcommand{\ep}{\,\epsilon}

\newcommand{\Dbar}{\,\overline{D}}
\newcommand{\qbar}{\,\overline{q}}

\newcommand{\fdr}{\,[\de^4q]}

\newcommand{\dr}{\,\frac{\de^nq}{\mu_R^{\ep}}}
\newcommand{\qbarslash}{\,\overline{\slashed{q}}}

\newcommand{\amp}{\,\mathcal{M}}
\newcommand{\bqa}{\begin{align}}
\newcommand{\eqa}{\end{align}}
\newcommand{\nl}{\nonumber \\}
\newcommand{\be}{\,\begin{equation}}
\newcommand{\ee}{\,\end{equation}}
\newcommand{\bes}{\,\begin{equation*}}
\newcommand{\ees}{\,\end{equation*}}
\usepackage{axodraw4j}
\usepackage{pstricks}
\usepackage{color}

\woctitle{LHCP2013}
\begin{document}
\title{	The $\mathbf{{\gamma\gamma}}$ decay of the Higgs boson in FDR }

\author{	Alice M. Donati 
			\inst{1}
			\fnsep\thanks{\email{adonati@ugr.es}. Speaker at LHCP2013 (16/5/2013).} 
		\and
        	Roberto Pittau
			\inst{1}
			\fnsep\thanks{\email{pittau@ugr.es}} 
		}

\institute{	Departamento de F\'{\i}sica Te\'orica y del Cosmos and CAFPE,\\ Campus de Fuentenueva s.n., Universidad de Granada E-18071 Granada, Spain 
        }

\abstract{
We review the first complete calculation performed within the Four Dimensional Regularization scheme (FDR), the recently-proposed approach for addressing multi-loop calculations, which is simultaneously free of infinities and gauge-invariant in 4 dimensions. 
As a case study, the 1-loop-induced amplitude for the Higgs boson decay into two photons was calculated in arbitrary gauge. The result obtained, identical to that assessed with standard methods, stands as an explicit test of the gauge-invariance property of FDR. Moreover, the calculation provides an insight into the use of the technique, in particular in the presence of fermions. 
}
\maketitle
\section{Introduction}
\label{intro}
In absence of a strong signal of new physics at LHC~\cite{:2012gk,:2012gu}, precision physics provides one of our best opportunities of investigating the unknown, searched as a tiny deviation from the Standard Model (SM). 
This requires the computation of more and more involved Radiative Corrections (RC), which is very demanding from a technical point of view. While a significant breakthrough has been made in dealing with multi-leg processes at 1-loop ~\cite{Ossola:2006us,Berger:2008sj,Giele:2008ve,Ellis:2011cr}, little simplification has been achieved in the context of multi-loop calculations~\cite{Mastrolia:2012an,Mastrolia:2011pr,Badger:2012dv,Johansson:2012zv,Kleiss:2012yv}.
In the usual framework, Dimensional Regularization (DR)~\cite{'tHooft:1972fi}, infinities arise at the intermediate steps of the calculation, forcing a huge analytic work in order to check all needed cancellations,  before even starting to calculate the physically-relevant contribution. This has pushed the quest of alternative approaches in 4 dimensions~\cite{Freedman:1991tk,delAguila:1997kw,delAguila:1998nd,Battistel:1998sz,Cherchiglia:2010yd,Wu:2003dd}. 
In this context, Four Dimensional Regularization was proposed \cite{Pittau:2012zd} as a method which is free of infinities, 4-dimensional and gauge-invariant at the same time. There are obvious advantages following from these characteristics: finiteness means that no renormalization is required, i.e. no counter-terms must be added to the Lagrangian; a fixed number of dimensions opens up the option of fully exploiting numerical integration; finally gauge-invariance provides a tool for testing the results, as well as guaranteeing that the correct expression for the amplitude, included the exact rational term, is straightforwardly obtained. 
In FDR this is all achieved via a simple re-interpretation of the loop integral, which is defined in such a way that ultraviolet (UV) infinities do not occur, at the price of introducing an arbitrary scale $\mu$ playing the role of the renormalization scale. The procedure works because the FDR construcion respects gauge-invariance. 

\subsection*{$\mathbf{H\rightarrow \gamma \gamma}$: a case study}

\begin{figure}
\centering
  \begin{picture}(177,51) (79,-77)
    \SetWidth{1.0}
    \SetColor{Black}
    \Line[dash,dashsize=4.4](80,-51)(103,-51)
    \Line[arrow,arrowpos=0.5,arrowlength=5,arrowwidth=2,arrowinset=0.2](103,-51)(133,-68)
    \Line[arrow,arrowpos=0.5,arrowlength=5,arrowwidth=2,arrowinset=0.2](133,-69)(133,-34)
    \Line[arrow,arrowpos=0.5,arrowlength=5,arrowwidth=2,arrowinset=0.2](133,-33)(103,-50)
    \Photon(133,-33)(156,-27){1.5}{2}
    \Photon(133,-69)(156,-75){1.5}{2}
    \Line[dash,dashsize=4.5](180,-51)(203,-51)
    \Photon(232,-33)(255,-27){1.5}{2}
    \Photon(232,-69)(255,-75){1.5}{2}
    \Photon(202,-51)(232,-34){1.5}{3}
    \Photon(202,-51)(232,-68){1.5}{3}
    \Photon(232,-33)(232,-68){1.5}{3}
  \end{picture}
\caption{Example of 1-loop diagrams contributing to the amplitude for the Higgs decay into two photons; the process can be mediated by either a fermionic loop or a vectorial one. }
\label{diagrams}       
\end{figure}
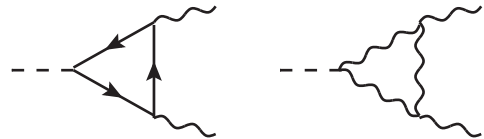

In a recent work \cite{Donati:2013}, the first application of FDR to a complete calculation in a realistic theory was presented: the 1-loop on-shell amplitude for the Higgs decay into two photons was calculated in arbitrary $R_{\xi}$-gauge, thereby explicitly verifying that the method respects gauge invariance. Due to its relevance and simplicity, this result, known since a long time~\cite{Ellis:1975ap,Ioffe:1976sd,Shifman:1979eb,Rizzo:1979mf}, has been recently reconsidered in several studies~\cite{Cherchiglia:2012zp,Shao:2011wx,Dedes:2012hf,Piccinini:2011az,Jegerlehner:2011jm,Huang:2011yf,Shifman:2011ri,Gastmans:2011wh,Gastmans:2011ks,Bursa:2011aa,Marciano:2011gm}. 
Since no $H\gamma\gamma$ interaction is described by the SM Lagrangian,  1-loop diagrams like those of fig.~\ref{diagrams}, provide the leading order contribution to the amplitude, which is therefore finite; however, because  divergences arise at intermediate steps, it is necessary to work within a regularization framework. 
The process is  mediated by either a fermionic or a bosonic loop, which contribute separately to the amplitude. 
By denoting with $k_1$ and $k_2$ the momenta of the photons, 
$\amp = \amp^{\mu\nu}\,
		\varepsilon^*_{\mu}(k_1)\,\varepsilon^*_{\nu}(k_2)$, and
\be \label{eq_amp_form_factors}
	\amp^{\mu\nu}(\beta,\eta) = 	
	 	\Big( 
			\widetilde{\amp}_W(\beta)
			+\sum_f N_c Q^2_f \,\widetilde{\amp}_f(\eta)
		\Big)
		\;{T}^{\mu\nu}\;.
\ee
$\widetilde{\amp}_W$ and $\widetilde{\amp}_f$ are scalar form factors of mass dimension $-1$, $\beta$ and $\eta$ are dimensionless kinematic parameters defined as
\be
	\beta = \frac{4\,M_W^2}{M_H^2}\,, \qquad 
	\eta  = \frac{4\,m_f^2}{M_H^2}\,,
\ee
and the tensorial structure is dictated by onshellness and gauge invariance to be
${T}^{\mu\nu} = k_1^{\nu} k_2^{\mu} -(k_1\cdot k_2) \; g^{\mu\nu}$ . 
The explicit expressions for the form factors, calculated entirely in the FDR framework \cite{Donati:2013}, were found consistent with the standard result (for example~\cite{Marciano:2011gm}):
\begin{align}
	\widetilde{\amp}_W(\beta) & =
		\frac{i\,e^3}{(4\pi)^2s_W M_W}\;
		\Big[\, 
			2 + 3\beta + 3 \beta (2-\beta) f(\beta)
		\,\Big]\,,
	\label{eq_W_form_factor}
	\\
	\widetilde{\amp}_f(\eta) & =
		\frac{-i\,e^3}{(4\pi)^2s_W M_W}\;
		2\eta\,\Big[\, 
			1 + (1-\eta) f(\eta)
		\,\Big]\,,
	\label{eq_f_form_factor}
\end{align}
where $s_W=\sin\theta_W$ is the sine of the Weinberg mixing angle, and
\footnote{$\varepsilon>0$ is a small imaginary part allowing for the analytic continuation of the result to any value of $x$.} 
\bes
	f(x) = 
	-\frac{1}{4}\ln^2
		\Big( 
			\tfrac{1+\sqrt{1-x+i\varepsilon}}{-1+\sqrt{1-x+i\varepsilon}}
		\Big)\,, \qquad  x = \frac{4\,M^2}{s}
\ees
is a parametrization of the scalar triangle

\begin{figure}[h!]
\centering
\begin{picture}(100,50) (55,-39)
    \SetWidth{1.0}
    \SetColor{Black}
    \Line(56,-14)(72,-14)
    \Line(72,-14)(96,2)
    \Line(96,2)(96,-30)
    \Line(96,-30)(72,-14)
    \Line[dash,dashsize=3.8](96,2)(112,10)
    \Line[dash,dashsize=3.8](96,-30)(112,-38)
    \Text(120,-20)[lb]{{\Black{
    		$= - \,\frac{2\,i\,\pi^2}{s}\, f(x)$}}}
  \end{picture}
\end{figure}
(solid lines denote momenta with mass $m$, dashed ones massless momenta).

This result will serve as a standpoint to illustrate some relevant features of the FDR scheme.
The outline of the contribution is as follows: in section~\ref{The_Method} the method is briefly reviewed; in section~\ref{Gauge_Invariance} the gauge-invariance property of the method is illustrated; finally in section~\ref{Fermions_in_FDR}, a technical remark on the treatment fermionic loops is made. 

\section{The FDR Method} \label{The_Method}

We will use the following notation : 
\be
	D_p = (q+p)^2-m_p^2\,, \qquad  d_q = m_p^2-p^2-2(q\cdot p)\,.
\ee
Consider, as the simplest example, 
\bes
	\frac{1}{D_0 \,D_p} 
\ees
which is obviously an UV-divergent integrand. Let us add a small arbitrary mass $\mu$ to the loop momentum, 
$q^2 \rightarrow \overline{q}^2 = q^2-\mu^2$, such that 
\bes
 	\lim_{\mu\rightarrow0}\frac{1}{\Dbar_0 \,\Dbar_p} = 
	\frac{1}{D_0 \,D_p}\,.
\ees
We can then use the partial fraction identities
\begin{align}
	\frac{1}{\Dbar_0} &= 
	\frac{1}{\qbar^2}\;\Bigg( 1+\frac{m_0^2}{\Dbar_0} \Bigg)\,,
	\\
	\frac{1}{\Dbar_p} &=
	\frac{1}{\qbar^2}\;\Bigg( 1+\frac{d_q}{\Dbar_p} \Bigg)\,
\end{align}
in order to expand the original integrand into a divergent term plus more and more convergent ones
\bes \label{expansion}
	\frac{1}{\Dbar_0 \Dbar_p}\,
	= 
	\Bigg[\frac{1}{\qbar^4}\Bigg]
			+\frac{d_p}{\qbar^4\Dbar_p}
			+\frac{m_0^2}{\qbar^2\Dbar_0\Dbar_p} \,.
\ees
The $\mu$-parametrization was essential in order to avoid spurious infrared (IR) divergences in the terms of the expansion. 
Notice that the term in squared brackets
- the only one to be UV-divergent - does not depend on any physical scale; the remaining part is finite and contains all the kinematical information. 
By means of the partial fraction identity we have decoupled the physical and the unphysical degrees of freedom of the integrand. The divergent term resembles a vacuum bubble (see fig.~\ref{vacuum_bubble}), universal and process-independent, and as such it should not be taken into account when calculating physical quantities.
Following this logic, the FDR integral is defined as the integral in 4 dimensions of the finite part alone:
\bes
	\int \fdr \frac{1}{D_0\,D_p} =
	\lim_{\mu\rightarrow0}
	\int \text{d}^4q \,
	\Bigg( \frac{d_p}{\qbar^4\Dbar_p}
			+\frac{d_0}{\qbar^2\Dbar_0\Dbar_p} \Bigg) \Bigg|_{\mu=\mu_R}
			\,.
\ees
This can be generalized to any Green's function.
To be more comprehensive, the symbol $\fdr$ means:
\begin{enumerate}
\item parametrizing in terms of $\mu$;
\item using the partial fraction identity in order to decouple the vacuum configurations (i.e. all integrands that only depend on the unphysical scale $\mu$);
\item dropping the vacuum configurations, or more precisely subtracting from the integrand the logarithmic infinities (an approach dubbed \emph{topological renormalization} in \cite{Pittau:2013b});
\item integrating the finite part in 4 dimensions;
\item taking the limit $\mu\rightarrow 0$, until a logarithmic divergence is met (more rigorously - performing a first order Taylor expansion around $\mu=0$); 
\item evaluating at $\mu = \mu_R$, where $\mu_R$ is the renormalization scale which separates the UV regime from the physical sector. The last point is equivalent to returning to the finite part of the integrand the IR component of the vacuum configuarations that had been naively dropped altogether with the divergent one. 
\end{enumerate}
The FDR integral has all of the good properties that we would like it to possess in order to perform calculations in Quantum Field Theory. First of all, it is just an integral, i.e. it is a linear operator and it is invariant under shift of integration momenta, which in particular implies invariance under momentum routing (this cannot be achieved by simply adding a cut-off or a regularizing distribution). Moreover, it is 4-dimensional, finite and independent of the UV regulator $\mu$. Finally it is gauge-invariant by construction. 

\begin{figure}
\centering
  \begin{picture}(90,60) (52,-19)
    \SetWidth{1.0}
    \SetColor{Black}
    \Arc(78,1)(19,90,450)
    \Vertex(59,1){2}
    \Text(107,-15)[lb]{\Large{\Black{$= \bigg[\frac{1}{\qbar^4}\bigg]$}}}
    \Text(53,17)[lb]{\Large{\Black{$\mu$}}}
  \end{picture}
\caption{1-loop topology of the universal divergent vacuum integrand. The dot means that the propagaor is squared.  }
\label{vacuum_bubble}       
\end{figure}
 
\section{Gauge Invariance} \label{Gauge_Invariance} 

Gauge-invariance is one of the key features of FDR, distinguishing it from other 4-dimensional methods. Consider, as an example, the bosonic contribution to the amplitude for the $H\rightarrow \gamma\gamma$ process, reported in eq.~\eqref{eq_W_form_factor}: in the viewpoint of gauge invariance this is interesting because of the presence of a rational term independent of the kinematics.   Indeed, it is obvious that terms conveying the kinematical dependence are equivalent in FDR and DR; however, a potential ambiguity remains in the constant term, because FDR and DR subract infinities at different stages of the calculation. In general, the rational term can be fixed by enforcing gauge-invariance \cite{Piccinini:2011az,Dedes:2012hf} or momentum-routing invariance \cite{Cherchiglia:2012zp} as extra constraints of the amplitude. Nevertheless FDR and DR alike automatically respect gauge invariance, thereby leading straightforwardly to the same correct constant.
Obtaining eq.~\eqref{eq_W_form_factor} in FDR is therefore an evidence of the gauge invariance property of the method, even more so because the calculation was performed in arbitrary $R_{\xi}$ gauge. 

What guarantees gauge invariance in FDR? The shift invariance of the FDR integral together with the \emph{global treatment} of $\mu^2$: if
every $q^2$ of the amplitude is promoted to its barred counter-part $q^2-\mu^2$, the usual simplifications between numerator and denominator can take place, thereby preserving gauge-invariance. For example, consider the trivial identity
\be \label{identity}
	\frac{q^2}{D^2} = \frac{1}{D}+\frac{m^2}{D^2} \;\leftrightarrow\;
	\frac{\overline{q}^2}{\overline{D}^2} = 
	\frac{1}{\overline{D}}
	+\frac{m^2}{\overline{D}^2}\,.
\ee
with $D=q^2-m^2$.
It still holds after subtracting $\mu^2$ everywhere, in the sense that
by integrating in the FDR fashion both sides of the second identity in eq.~\eqref{identity}  one obtains the same result, i.e.
\bes
	\int[\text{d}^4q]\, \frac{\overline{q}^2}{\overline{D}^2} = 
	\int[\text{d}^4q]\, \frac{1}{\overline{D}}
	+ \int[\text{d}^4q]\, \frac{m^2}{\overline{D}^2} \,.
\ees
This explains why a $q^2$ and its associated $\mu^2$ should never be treated separately. However, there are cases in which a $\mu^2$ does appear alone, for example when performing a tensorial reduction (Passarino-Veltman reduction \cite{Passarino:1978jh} in FDR is extensively treated in \cite{Donati:2013}); e.g.
\begin{align}
	\int\fdr \frac{q^{\mu}q^{\nu}}{\overline{D}^2} 
	& = 
	\frac{g^{\mu\nu}}{4} \int\fdr \; \frac{q^2}{\overline{D}^2}
	\nl
	& = 
	\frac{g^{\mu\nu}}{4} \int\fdr \;
	\Bigg( \,\frac{\overline{q}^2}{\overline{D}^2}
	+ \frac{\mu^2}{\overline{D}^2} \,\Bigg)\,. \nonumber
\end{align}
The $\mu^2$ in the last equation is a reminder of the tensorial structure of the original rank-2 integral; as such, it effectively plays the role of a $q^{\mu}q^{\nu}$ in the power-counting, as well as in the FDR expansion: 
\bes
	\frac{(q^{\mu}q^{\nu};\mu^2 )}{\Dbar^2} 
	= (q^{\mu}q^{\nu};\mu^2 ) \,\Bigg(\,
	\Bigg[ \frac{1}{\qbar^4} \Bigg]
	+ \frac{2m^2}{\qbar^6}
	+ \Bigg\{ \frac{2m^4}{\qbar^6\Dbar}+\frac{m^4}{\qbar^4\Dbar^2} \Bigg\}
	\Bigg) .
\ees
This means that this term is not killed in the limit $\mu\rightarrow 0$, rather
\bes
\int[\text{d}^4q] \frac{\mu^2}{\overline{D}^2}
= i \pi^2 m^2\,.
\ees
This type of contributions are essential to preserve gauge invariance, equivalently to DR, when a finite term is obtained as the product of an $O(\ep)$-term and a single pole $1/\ep$. More explicitly one can prove that 
\be
\label{correspondence}
	\int\dr \frac{(-\widetilde{q}^2)^k}{\Dbar^{(n)}_0\Dbar^{(n)}_1\ldots} 
	=
	\int [d^4q] \frac{(\mu^2)^k}{\Dbar_0\Dbar_1\ldots}\,,
\ee
where $\widetilde{q}^2 = (q^{(n)})^2-q^2$ is the $\ep$-dimensional part of an $n$-vector, and the superscript $(n)$ denotes an object living in $n$ dimensions. 

\section{Fermions in FDR} \label{Fermions_in_FDR}
A brief comment is due regarding the treatment of Dirac matrix strings in FDR. How should the Dirac propagator, 
\bes
	\frac{i}{\slashed{q}-m}=\frac{i\;\big(\,\slashed{q}+m \,\big)}{q^2-m^2}\,,
\ees
be parametrized in terms of $\mu$? The guideline is again that of preserving the usual simplifications between numerator and denominator, in order to ultimately respect gauge invariance. This can be achieved either by promoting $q^2\rightarrow\qbar^2$ after calculating the trace (in 4 dimensions), or by replacing $\slashed{q}\rightarrow \qbarslash \equiv q \pm \mu $ directly in the string \footnote{Thanks also to the fact that FDR integrals involving odd powers of $\mu$ vanish~\cite{Pittau:2012zd}.}.
In the latter case, $\qbarslash$ is defined according to its position within the string:
\bes \label{eq_fermion_prescription}
	( \, \qbarslash \; \gamma^{\alpha_1}\ldots\gamma^{\alpha_n}
		\qbarslash\, \ldots )
	= 
	( \, (\slashed{q}\pm\mu) \; \gamma^{\alpha_1}\ldots\gamma^{\alpha_n} (\slashed{q}\mp(-)^{n}\mu)  \ldots ) \,.
\ees
The sign of the first $\qbarslash$ is chosen arbitrarily; in the following $\qbarslash$, the sign is opposite if an even number of $\gamma$-matrices occur between the two $\qbarslash$'s, and it is the same in the case of an odd number of $\gamma$-matrices. And so on for all pairs of $\slashed{q}$'s occurring. \\
This prescription is explicitely verified by eq.~\eqref{eq_f_form_factor}, i.e. the fermionic contribution to the amplitude of \hbox{$H\rightarrow\gamma\gamma$}.

\section{Conclusions} \label{Conclusions}
The calculation of the amplitude for the Higgs decay into two photons stands as the first test of FDR in the electroweak theory, explicitely verifying - at 1-loop - that the method respects gauge invariance, and providing a detailed example of calculation \cite{Donati:2013}.  We have made it clear that FDR is equivalent to DR at 1-loop, in the sense of a 1-1 correspondence between Feynman diagrams and analytic expressions; the advantage of FDR lies in that there is no need of verifying massive cancellations of unphysical contributions, and - in the case of processes that demand renormalization - that the Lagrangian does not require the addition of counter-terms. Already at two loops, we expect  FDR and topological renormalization to prove significantly more convenient with respect to the usual approaches \cite{Pittau:2013b}, on the practical level of making calculations easier, and on the more theoretical viewpoint of dealing with non-renormalizable theories. 
As a matter of fact, the same mechanism that guarantees gauge invariance at 1-loop is believed to work unchanged in the case of more loops, as well as in the presence of IR and collinear divergences. Verifying these conjectures is the subject of present investigations. 

\bibliography{bibliography.bib}

\end{document}